# Ensemble Classifier Design Tuned to Dataset Characteristics for Network Intrusion Detection

Zeinab Zoghi and Gursel Serpen

Electrical Engineering & Computer Science, University of Toledo, Ohio, USA

*Abstract* – Machine Learning-based supervised approaches requires highly customized and fine-tuned methodologies to deliver outstanding performance. This paper presents a dataset-driven design and performance evaluation of a machine learning classifier for the network intrusion dataset UNSW-NB15. Analysis of the dataset suggests that it suffers from class imbalance and class overlap in the feature space. We employed ensemble methods using Balanced Bagging (BB), eXtreme Gradient Boosting (XGBoost), and Random Forest empowered by Hellinger Distance Decision Tree (RF-HDDT). BB and XGBoost are tuned, and Random Forest (RF) classifier is supplemented by the Hellinger metric to address the imbalance issue. Two new algorithms are proposed to address the class overlap issue in the dataset. These two algorithms are leveraged to help improve the performance of the testing dataset by modifying the final classification decision made by three base classifiers as part of the ensemble classifier which employs a majority vote combiner. The proposed design is evaluated for both the binary and multi-category classification. Comparing the proposed model to those reported on the same dataset in the literature demonstrate that the proposed model outperforms others by a significant margin in binary and multi-category classification.

# 1. Introduction

A careful analysis of a dataset for classifier development through machine learning approaches is essential to identify the inherent issues in that dataset and consequently to address them to obtain the best performance possible [1]. We presented an analysis on the UNSW-NB15 dataset in our earlier study [2]. The analysis showed that the dataset possesses two issues that need to be addressed, namely class imbalance [3, 4] and class overlap [5]. Additionally, it was determined that the class imbalance for this dataset is composed of both between-class and within-class imbalance.

Between-class imbalance corresponds to the case where one class or multiple classes in a dataset are underrepresented in comparison with other classes. In other words, a dataset shows significantly unequal distribution among its classes. For the UNSW-NB15 dataset, normal records constitute 87% of all records while the combined record count for all 9 attack classes is only 13%. Additionally, 13% of records in the overall dataset are not equally distributed among the 9 attack classes as 65% of all attack records belong to the Generic attack class while only 0.0008% of all attack records belong to the Worms attack class.

Within-class imbalance, on the other hand, represents the case where one class is comprised of several different subclasses with different distributions. To discover whether the classes are made up of imbalanced sub-clusters, we employed two visualization techniques, namely Principal Component Analysis (PCA) and t-distributed Stochastic Neighbor Embedding (t-SNE) [6]. Consequent analysis exposed the within-class imbalance in the UNSW-NB15 dataset. As an example, the Exploits attack class is composed of several different size clusters where clusters are compact. The clustering is different for the Worms attack class for which there are two but different size cluster formations. The larger cluster grouping is composed of multiple subgroupings with gaps in between. All classes have multiple clusters of different sizes and spread across the two-dimensional analysis space. Many classes are composed of a few relatively large clusters and many small clusters.

The boundaries separating classes are not clear cut: there is noticeable overlap between or among multiple clusters belonging to different classes, which indicates that this dataset also suffers from the so-called "class overlap problem." In particular and as a prominent case, many attack class records mimic the behavior of the Normal records. While the emphasis of intrusion detection systems is detecting and identifying the malicious network traffic, if it is trained by this dataset without addressing the overlap problem, a satisfactory outcome may not be achieved.

In this paper, an improved ensemble supervised [7] machine learning classifier is proposed and its performance is evaluated using the UNSW-NB15 dataset. The model combines the decisions from namely BB, XGBoost, and RF-HDDT to addresses imbalance problem. According to the previously reported studies in the literature [8-18], the tree-based algorithms outperformed other commonly used ML algorithms such as Support Vector Machine (SVM) and K-Nearest Neighbor (KNN) for the UNSW-NB15 dataset. In light of this finding, this study also employs tree-based classification algorithms, XGBoost and RF, in addition to BB within the ensemble framework. The hyper-parameters of BB [19, 20] and XGBoost [21] classifiers are tuned to handle the imbalanced data. For the RF [22] classifier, the Hellinger distance [23, 24] is utilized to measure the similarity between two probability distributions for unequal distributed classes. Two novel algorithms are also proposed to handle the class overlap issue in the dataset by measuring the prediction error made by the ensemble classifier during the validation phase and revising the final classification decision in the testing phase. A min-max scaler is implemented on the dataset in the preprocessing phase followed by two wrapper feature selectors [25], namely Elastic Net and SFS. Subsequent sections first present the UNSW-15NB dataset followed by preprocessing and proposed algorithms to customize the ensemble classifier design to the characteristics of the UNSW-NB15 dataset.

## 2. Dataset

The UNSW-NB15 network security dataset had been released in 2015 by Moustafa et al. [26, 27]. This dataset is comprised of 2,540,044 records, available through the UNSW webpage [28] and contains realistic, modern normal and malicious network activities. The structure of this dataset is more complex in comparison with the other benchmark datasets [29-33], which makes the UNSW-NB15 more comprehensive for evaluating the network intrusion detection systems in a more reliable way [30].

The records for the UNSW-NB15 dataset were gathered by the IXIA traffic generator [34] using three virtual servers. Two servers were configured to distribute the normal traffic packets and the third one was configured to spread the abnormal traffic packets. A total of 49 features including those that are packet-based and flow-based were extracted from the records by Argus [35] and Bro-IDS [36] tools. Packet-based features are extracted from a packet header and its payload (also called packet data). In contrast, flow-based features are produced from the sequence of packets, from a source to a destination, traveling in the network. The features are categorized into basic, content and time. Others are labeled as general-purpose features and connection features. General-purpose features include those which are intended to explain the purpose of an individual record while connection features capture the characteristics of the connection between a hundred sequentially-ordered records. The last two features represent attack categories and labels.

Attacks are categorized as Analysis, Backdoor, DoS, Exploits, Fuzzers, Generic, Reconnaissance, Shellcode, and Worms. The Normal class contains 2,218,761 records while Fuzzers, Analysis, Backdoors, DoS, Exploits, Generic, Reconnaissance, Shellcode, and Worms signatures include 24246, 2677, 2329, 16535, 44525, 215481, 13987, 1511, and 174 records, respectively. The dataset is subsampled and split into training and testing subsets. The training split contains 175,341 records and the test split contains 82,332 records each of which contains the records belonging to 9 different attack classes as detailed in Table 1.

**Table 1** Number of records in training and testing subsets for each class

| Classes | Training Subset | Testing Subset | All |
|---|---:|---:|---:|
| Normal | 56,000 | 37,000 | 93,000 |
| Analysis | 2,000 | 677 | 2,677 |
| Backdoor | 1,746 | 583 | 2,329 |
| DoS | 12,264 | 4,089 | 16,353 |
| Exploits | 33,393 | 11,132 | 44,525 |
| Fuzzers | 18,184 | 6,062 | 24,246 |
| Generic | 40,000 | 18,871 | 58,871 |
| Reconnaissance | 10,491 | 3,496 | 13,987 |
| Shellcode | 1,133 | 378 | 1,511 |
| Worms | 130 | 44 | 174 |
| **Total Number of Records** | **175,341** | **82,332** | **257,673** |

The class distribution in UNSW-NB15 is shown in Fig1. The skewed class proportion, represents the between-class imbalance. In between-class imbalance, although the distribution of the sub-classes within minority classes is even as is within majority classes, the sample size between the majority (Normal) and minority (Worms) classes defers.

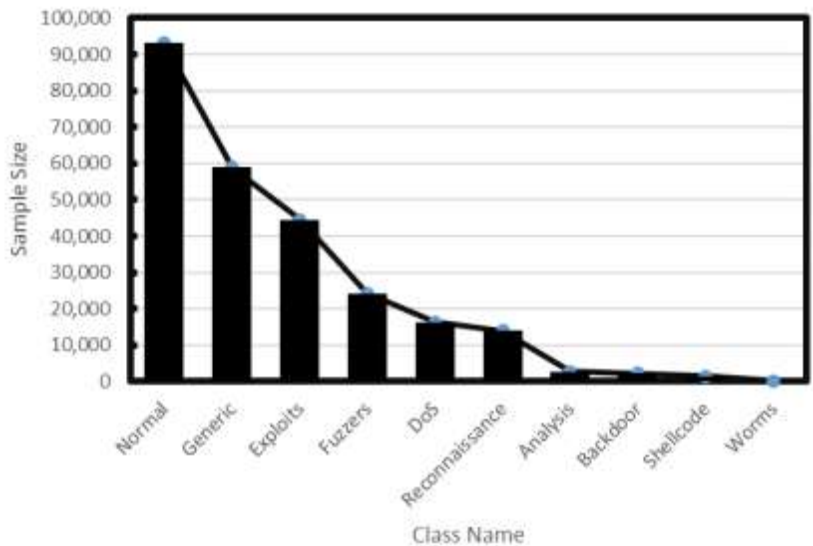

Fig1. UNSW-NB15 Class Distribution

To measure how severe the class imbalance is, Imbalance Ratio (IR: majority class distribution/minority class distribution) is calculated on the UNSW-NB15 samples. Since the UNSW-NB15 is a multi-class dataset, it is decomposed to $m(m-1)/2$ binary-class subsets using One-Versus-One (OVO) method, where $m$ is the number of classes. Then, the IR is computed for each class against the rest. According to Fernández, Alberto, et al. [37], any IR value greater than 1.5 is considered as imbalanced.

The occurrence of the classes, empirical class distribution, and the IR is shown in Table 2. The greatest and the smallest IR value is 514.29 and 1.50, respectively, that indicates the Normal and Worms possess the most, and Fuzzers and DoS possess the less imbalance issue. Since the imbalance threshold is 1.5, there is no imbalance issue between Backdoor/Analysis, Generic/Exploits, Reconnaissance/DoS, Shellcode/Analysis, and Shellcode/Backdoor.

**Table 2** Number of records in training and testing subsets for each class

| Occurrence | | | | | Empirical Class Distribution | | | | |
|---|---|---|---|---|---|---|---|---|---|
| 93,000/2,677/2,329/16,353/44,525/24,246/58,871/13,987/1,511/174 | | | | | 0.36/0.01/0.01/0.06/0.17/0.09/0.23/0.05/0.01/0.0007 | | | | |
| | | | | | IR | | | | |
| Class Name | Normal | Analysis | Backdoor | DoS | Exploits | Fuzzers | Generic | Recon | Shellcode | Worms |
| Normal | 1 | 36.00 | 36.00 | 6.00 | 2.12 | 4.00 | 1.57 | 7.20 | 36.00 | 514.29 |
| Analysis | 36.00 | 1 | 1 | 6.00 | 17.00 | 9.00 | 23.00 | 5 | 1 | 14.29 |
| Backdoor | 36.00 | 1 | 1 | 6 | 17 | 9 | 23 | 5 | 1 | 14.29 |
| DoS | 6.00 | 6.00 | 6.00 | 1 | 2.83 | 1.50 | 3.83 | 1.2 | 6 | 85.71 |
| Exploits | 2.12 | 17 | 17 | 2.83 | 1 | 1.89 | 1.35 | 3.4 | 17 | 242.86 |
| Fuzzers | 4.00 | 9 | 9 | 1.50 | 1.89 | 1 | 2.56 | 1.80 | 9.00 | 128.57 |
| Generic | 1.57 | 23 | 23 | 3.83 | 1.35 | 2.56 | 1 | 4.60 | 23 | 328.57 |
| Reconnaissance | 7.20 | 5 | 5 | 1.2 | 3.4 | 1.80 | 4.60 | 1 | 5 | 71.43 |
| Shellcode | 36.00 | 1 | 1 | 6 | 17 | 9.00 | 2 | 5 | 1 | 14.29 |
| Worms | 514.29 | 14.29 | 14.29 | 85.71 | 242.86 | 129.57 | 328.57 | 71.43 | 14.29 | 1 |

The high IR values in Table 2 between most of the classes in UNSW-NB15 indicate that the imbalance issue in this dataset is severe and needs to be handled. According to [2], there is also overlapping issue in this dataset. So, the class imbalance issue becomes more serious when class overlap is also present in the dataset. To address both issues, two novel algorithms are proposed in the following sctions.

## 3. Data Preprocessing and Exploratory Classifier Evaluation

In this section, several classifiers are evaluated and those with high performance were picked to form the proposed model. Once the top classifiers were selected, a multistep preprocessing approach was implemented on UNSW-NB15 to improve the performance of the classifiers by preparing the data for training them effectively. However, the first preprocessing step was done preceding the classifier evaluation. In this step, source IP address, source port number, destination IP address, destination port number, record_start_time, and record_last_time features were removed; the features that potentially did not help with the prospect of developing a classifier that would be applicable in a general sense [38-42].

SVM, naïve Bayes (NB), Bagging, multi-layer perceptron neural network (MLP-NN), RF, Extremely Randomized Trees (ERT), AdaBoost, Gradient Tree (GT), BB, XGBoost, and Easy Ensemble (EE) were employed on the dataset with the 42 remaining features in order to identify high-performing base classifiers for an ensemble design. The results are shown in Table 2 in terms of missed alarm rate.

**Table 2** Missed alarm rate values for the set of 11 classifiers

| Classes | SVM | NB | Bagging | MLP-NN | RF | ERT | AdaBoost | GT | BB | XGBoost | EE |
|---|---|---|---|---|---|---|---|---|---|---|---|
| Analysis | 1.00 | 1.00 | 0.99 | 1.00 | 0.86 | 1.00 | 0.87 | 1.00 | **0.77** | 0.87 | 0.86 |
| Backdoor | 1.00 | 1.00 | 0.93 | 0.98 | 0.76 | 0.95 | 0.93 | 0.95 | **0.59** | 0.64 | 0.60 |
| DoS | 0.90 | 0.99 | 0.88 | 0.98 | 0.87 | 0.87 | 0.99 | 0.93 | **0.81** | 0.83 | 0.99 |
| Exploits | 0.90 | 0.97 | 0.21 | 0.82 | 0.21 | 0.28 | 0.70 | 0.09 | 0.42 | **0.04** | 0.99 |
| Fuzzers | 0.94 | 0.63 | 0.42 | 0.89 | 0.39 | 0.42 | 0.93 | 0.55 | 0.32 | **0.29** | 0.75 |
| Generic | 0.51 | 0.03 | 0.03 | 0.03 | 0.03 | 0.03 | 0.42 | 0.03 | 0.04 | **0.02** | 0.42 |
| Normal | 0.04 | 0.65 | 0.24 | 0.24 | **0.21** | 0.23 | 0.86 | 0.34 | 0.34 | 0.39 | 0.70 |
| Recon | 0.65 | 0.71 | 0.19 | 1.00 | **0.17** | 0.22 | **0.17** | 0.21 | **0.17** | 0.18 | 0.99 |
| Shellcode | 0.96 | 0.98 | 0.31 | 1.00 | 0.30 | 0.52 | 0.92 | 0.60 | **0.06** | 0.12 | 0.81 |
| Worms | 1.00 | 0.95 | 0.86 | 1.00 | **0.04** | 0.86 | 1.00 | 0.56 | 0.09 | 0.43 | 0.88 |

The results reveal that BB, XGBoost, and RF lead in their performances. The top classifiers were implemented along with Normalizer, Min-Max scaler, Robust scaler, Standard scaler, Quantile transformer, and Power transformer to identify the scalers, those increase the power of the classifiers. The results are presented in Table 3 in terms of accuracy. The first row represents the accuracy scores where the classifiers were implemented on the original dataset. The rest of the table contains the accuracy scores delivered by the classifiers implemented on the scaled dataset. Comparing the first line with the rest, indicates that the Min-Max scaler improved the performance of the classifiers.

**Table 3** Accuracy comparison among the performances of six different scalers

|  | BB (%) | XGBoost (%) | RF (%) |
|---|---|---|---|
| Original Dataset | 87.56 | 94.11 | 92.81 |
| Normalizer | 81.45 | 85.02 | 85.78 |
| Min-Max Scaler | **88.91** | **96.40** | **93.33** |
| Robust Scaler | 87.61 | 94.55 | 93.12 |
| Standard Scaler | 86.64 | 92.73 | 91.99 |
| Quantile Transformer | 87.32 | 93.95 | 93.12 |
| Power Transformer | 86.11 | 92.00 | 91.27 |

In the final preprocessing step, the redundant features from the remaining 42 features were removed. The high-performance classifiers were evaluated on UNSW-NB15 chi-squared, genetic algorithm, mRMR, relief, RFE, SFS. Utilizing Elastic Net and SFS algorithms increased the performance of classifiers in all cases shown in Figure 1. The Elastic Net feature selection algorithm identified 24 features which were then further processed by the SFS feature selection algorithm to reduce the selection to 8 features as listed in Table 4. The features marked by the 'X' are selected by the corresponding feature selection method in the same table. The first column represents 24 features selected by Elastic Net while the second column shows the 8 features from 24 as by the SFS algorithm. BB and XGBoost performed better with 24 features selected by the Elastic Net while RF performed better with only 8 features selected by the SFS as presented in Figure 1.

**Table 4** List of 24 features selected by Elastic Net and 8 features selected by SFS

| Feature Name | Elastic Net | SFS | Description |
|---|---|---|---|
| dur | X |  | Record total duration |
| proto | X | X | Transaction protocol |
| service | X | X | Contains the network services |
| state | X |  | Contains the state and its dependent protocol |
| spkts | X |  | Source to destination packet count |
| dpkts | X |  | Destination to source packet count |
| sbytes | X | X | Source to destination transaction bytes |
| dbytes | X |  | Destination to source transaction bytes |

| Feature Name | Elastic Net | SFS | Description |
|---|---|---|---|
| rate | X | X | Ethernet data rates transmitted and received |
| sttl | X | | Source to destination time to live value |
| dttl | X | | Destination to source time to live value |
| sload | X | | Source bits per second |
| dload | X | X | Destination bits per second |
| sloss | X | | Source packets retransmitted or dropped |
| dloss | X | | Destination packets retransmitted or dropped |
| sinpkt | X | | Source interpacket arrival time (mSec) |
| dinpkt | X | | Destination interpacket arrival time (mSec) |
| sjit | X | X | Source jitter (mSec) |
| djit | X | X | Destination jitter (mSec) |
| swin | X | | Source TCP window advertisement value |
| stcpb | X | | Destination TCP window advertisement value |
| dtcpb | X | | Destination TCP base sequence number |
| dwin | X | | Destination TCP window advertisement value |
| tcprtt | X | X | TCP connection setup round-trip time |

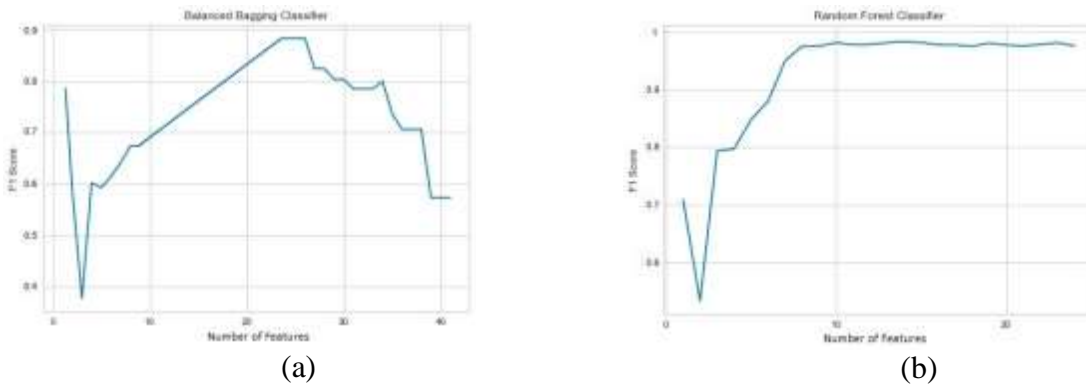

(a)            (b)

**Figure 1** (a) The F1-score of BB classifier across the number of input features. (b) The F1-score of RF classifier across the number of input features

## 4. Classifier development methodology

To deal with the imbalanced data, we utilized the BB and the XGBoost ensemble classifiers which offer a set of hyper-parameters that, through judicious adjustments of the same, help contribute to improved performance in the presence of the imbalanced data. To address the class overlap problem, we proposed two new algorithms, namely Algorithms #1 and #2 which are presented in Figures 2 and 5, respectively, and utilized them to process the classification outputs from the BB, and the XGBoost classifiers. Also, the performance of RF on the UNSW-NB15 dataset was somewhat lacking particularly with respect to the missed alarm rate metric. Decision Trees suffer from overfitting, axis-parallel splitting, and skewness sensitivity. The overfitting problem is mitigated by tree pruning, while axis-parallel splitting can be addressed by building a forest of orthogonal [22] and oblique decision trees [43]. Skewness sensitivity of decision trees arises due to utilizing popular splitting criteria including information gain and Gini measure. Hellinger's measure can address this problem due to its skew insensitivity property. Therefore, the Hellinger distance is leveraged as the split criterion in the construction of trees rather than the entropy value which is used as the default metric [24, 44] in the RF. Outputs of three classifiers, namely RF, BB and XGBoost, were provided as inputs to a combiner that implemented majority voting to determine the final class membership of an input data record under test.

## 4.1. Training

The training data set is split into two subsets: one subset, namely the training subset, and the second subset, namely the validation subset using stratified sampling [45]: 80% of the training set is extracted in order to train the proposed model and the remaining 20% is used for calculating the errors arising from the wrong predictions by the BB and XGBoost classifiers. The error calculation is made by the Algorithm #1 which will be explained in detail in the following paragraphs.

Each stratum is formed by ten different classes following a frequency-based distribution. In other words, the samples are picked randomly from each attack class as well as Normal records. Next, the training subset is processed with the Elastic Net feature selection algorithm. This algorithm selects 24 features out of 42 before the training subset is used to train the BB and XGBoost classifiers. Concurrently, the SFS algorithm selects the 8 most informative features out of 24 that were already selected among the original 42 by the Elastic Net. The output of SFS, which entails 8 features, is used to train the RF classifier.

Each classifier generates a matrix consisting of probability scores. The probability score of an input sample for BB is calculated as the mean class probabilities of trees. Also, the XGBoost algorithm computes the sum of outputs from the leaves of each individual tree, aka margin score. Then the sigmoid function maps the entire margin scores into the range between 0 and 1 that can be interpreted as a probability score. The first seven rows of the matrix produced by the BB are shown in Table 5. It consists of 10 columns, representing 10 (9 attack plus Normal) classes, and $N$ rows, where $N$ designates the number of data samples in the validation subset. To calculate the prediction errors, the validation set is input to the BB and XGBoost classifiers. Since the validation subset has 35,068 records representing 20% of the training set, each model generates a probability matrix consisting of 35,068 rows. Both the probability matrix and the confusion matrix produced by XGBoost and BB classifiers are used as the inputs of Algorithm #1 presented in Figure 2. This algorithm is employed to process the XGBoost and BB outputs to compute the prediction errors caused by the overlap of classes in the feature space which is inherent to the dataset. Algorithm #1 records the smallest membership score errors made by BB and XGBoost as a result of these classifiers incorrectly estimating the likelihood of the training samples being the member of a specific class type. Algorithm #1 then calculates the mean and standard deviation of the errors. These computed values for mean and standard deviation are next utilized by Algorithm #2 presented in Figure 5 to minimize the membership errors that may occur during the testing phase.

**Table 5.** First seven rows of a probability score matrix generated by BB classifier

| Row Index | Analysis | Backdoor | DoS | Exploit | Fuzzers | Generic | Normal | Recon | Shellcode | Worms |
|---|---|---|---|---|---|---|---|---|---|---|
| 0 | 0.0083 | 0.0103 | 0.0988 | 0.1506 | 0.1718 | 0.0167 | 0.0750 | 0.0212 | 0.4463 | 0.0003 |
| 1 | 0.0034 | 0.0044 | 0.0632 | 0.1580 | 0.4279 | 0.0240 | 0.2884 | 0.0118 | 0.0144 | 0.0045 |
| 2 | 0.0057 | 0.0087 | 0.0641 | 0.1094 | 0.5057 | 0.0141 | 0.2659 | 0.0143 | 0.0111 | 0.0011 |
| 3 | 0.0062 | 0.0101 | 0.0627 | 0.1301 | 0.5491 | 0.0118 | 0.1968 | 0.0154 | 0.0159 | 0.0017 |
| 4 | 0.0018 | 0.0036 | 0.0339 | 0.0843 | 0.6023 | 0.0113 | 0.2427 | 0.0079 | 0.0109 | 0.0013 |
| 5 | 0.0044 | 0.0080 | 0.0827 | 0.1667 | 0.5055 | 0.0165 | 0.1213 | 0.0149 | 0.0774 | 0.0027 |
| 6 | 0.0034 | 0.0051 | 0.0605 | 0.1625 | 0.4867 | 0.0242 | 0.2271 | 0.0105 | 0.0140 | 0.0061 |

The output of Algorithm #1 is a nested hashmap data structure with elements holding unordered lists, which constitutes Level-1 as shown in Figure 3, and consists of 10 items representing all attack classes along with the Normal class. Each item in Level-1 points to 9 sub-items in Level-2. The corresponding sub-items in Level-2 for each item will not hold the item itself in Level-1. For each sub-item in Level-2 of the nested list, there is a corresponding two-element list in Level-3. For further clarification, we can consider this sole nested list as two two-dimensional arrays with the same dimensionality as the confusion matrix (10 by 10) storing mean and standard deviation values. In other words, one matrix could hold the mean values, and another would hold the standard deviation values. Each row and column represent nine different attack classes along with the Normal class with the same order, similar to the rows and columns of the confusion matrix. These arrays store zeros along their main diagonal. The reason for that is that the aim of Algorithm #1 is to compute the prediction errors or those errors existing in the membership scores. Since the main diagonal is holding true positives in a confusion matrix, there is no error to calculate. That is the reason these entries are not represented in the nested list.

Algorithm #1 employs the confusion matrix: if a member of class A is misclassified $n$ different times as a member of class B, then this algorithm iterates $n$ times to calculate the difference between the probability scores of class B and class A as well as class B and the eight other classes. If the former difference is smaller than those of the other eight differences, this difference value (between class B and class A) is stored in a temporary variable (array DL) for further processing. Otherwise, the value is discarded. In the next step, the mean and standard deviation of the stored values are calculated and kept in arrays OLM and OLSD, respectively. These mean and standard deviation values are placed at the third level of the nested list, which constitute also the two outputs of Algorithm #1, where class A is the element of the first level of the list and class B is the element of the second level of the list.

Algorithm #1 – Compute Means and Standard Deviations

Mean-Standard-Deviation(CM, PS)
  in:
    two-dimensional array $CM_{10 \times 10}$ holding the confusion matrix
    two-dimensional array $PS_{n \times 10}$ holding the membership scores with n = the number of samples of validation subset
  out:
    two-dimensional array $OLM_{10 \times 10}$ initialized with zero
    two-dimensional array $OLSD_{10 \times 10}$ initialized with zero
  local:
    empty array DL representing the minimum value of the membership scores difference
    empty variable σ representing the computed Standard Deviation value
    empty variable μ representing the computed Mean value
    empty variables DT and D hold the values obtained by subtracting the membership scores
  constant:
    array AL = {*Analysis, Backdoor, DoS, Exploits, Fuzzers, Generic, Normal, Recon, Shellcode, Worms*}

```
1   for all x ∈ AL do                                ▷ AL is a list
2     for all y ∈ (AL − x) do
3       if CM_{x,y} > 0 then
4         j ← 0
5         for i ← 1 … CM_{x,y} do
6           D ← PS_{i,y} − PS_{i,x}                  ▷ where x is misclassified as y
7           for all z ∈ (AL − {x, y}) do
8             DT ← PS_{i,y} − PS_{i,z}
9             if DT < D then
10              count ← 1
11            end if
12          end for
13          if count = 0 then
14            DL_j ← D
15            j ← j + 1
16          end if
17        end for
18        sum ← 0 and N ← length(DL)
19        for h ← 1 … length(DL) do
20          sum ← sum + DL_h
21        end for
22        μ ← sum / N                                ▷ where N indicates the number of elements in DL
23        σ ← √((1/N) Σ_{i=1}^{N} (DL_i − μ)^2)
24        OLM_{x,y} ← μ
25        OLSD_{x,y} ← σ
26      end if
27    end for
28  end for
29  return OLM, OLSD
```

**Figure 2**. Pseudocode for Algorithm #1

To clarify how Algorithm #1 works, we next present a step-by-step trace of its application on the Analysis and the Backdoor attack classes. The first row of Table 6 (a) represents a confusion matrix for the Analysis attack, and Table 6 (b) represents the probability scores generated by the BB classifier in a two-dimensional array or matrix form after it is trained and its performance evaluated on the validation subset. Values of these matrices are held by CM and PS, two-dimensional array variables in the pseudocode of Algorithm #1, respectively. Initially, DL, OLM, and OLSD are empty lists and eventually holding values for distance, mean ($\mu$), and standard deviation ($\sigma$) values, respectively. AL is another list that contains the Normal and all nine attack classes as its elements in some arbitrary order.

**Table 6.** (a) Confusion matrix and (b) First two rows of probability scores matrix.

| | Analysis | Backdoor | DoS | Exploits | Fuzzers | Generic | Normal | Recon | Shellcode | Worms |
|---|---|---|---|---|---|---|---|---|---|---|
| Analysis | 501 | 124 | 0 | 0 | 46 | 0 | 3 | 3 | 0 | 0 |
| Backdoor | 0 | 302 | 0 | 0 | 0 | 12 | 0 | 0 | 0 | 1 |
| DoS | 3 | 0 | 270 | 4 | 0 | 0 | 0 | 0 | 5 | 0 |
| Exploits | 0 | 0 | 0 | 254 | 8 | 0 | 50 | 0 | 0 | 0 |
| Fuzzers | 3 | 0 | 0 | 32 | 345 | 0 | 23 | 0 | 0 | 0 |
| Generic | 1 | 0 | 0 | 0 | 9 | 138 | 0 | 0 | 0 | 0 |
| Normal | 0 | 0 | 0 | 54 | 78 | 0 | 876 | 0 | 0 | 0 |
| Recon | 0 | 0 | 0 | 0 | 0 | 0 | 8 | 132 | 0 | 0 |
| Shellcode | 0 | 0 | 0 | 17 | 0 | 0 | 0 | 1 | 187 | 0 |
| Worms | 1 | 0 | 0 | 1 | 0 | 0 | 0 | 0 | 2 | 74 |

(a)

| Index | Analysis | Backdoor | DoS | Exploits | Fuzzers | Generic | Normal | Recon | Shellcode | Worms |
|---|---|---|---|---|---|---|---|---|---|---|
| 1 | 0.78 | 0.81 | 0.02 | 0.01 | 0.24 | 0.08 | 0.11 | 0.19 | 0.08 | 0.22 |
| 2 | 0.09 | 0.54 | 0.01 | 0.03 | 0.41 | 0.04 | 0.01 | 0.06 | 0.12 | 0.14 |

(b)

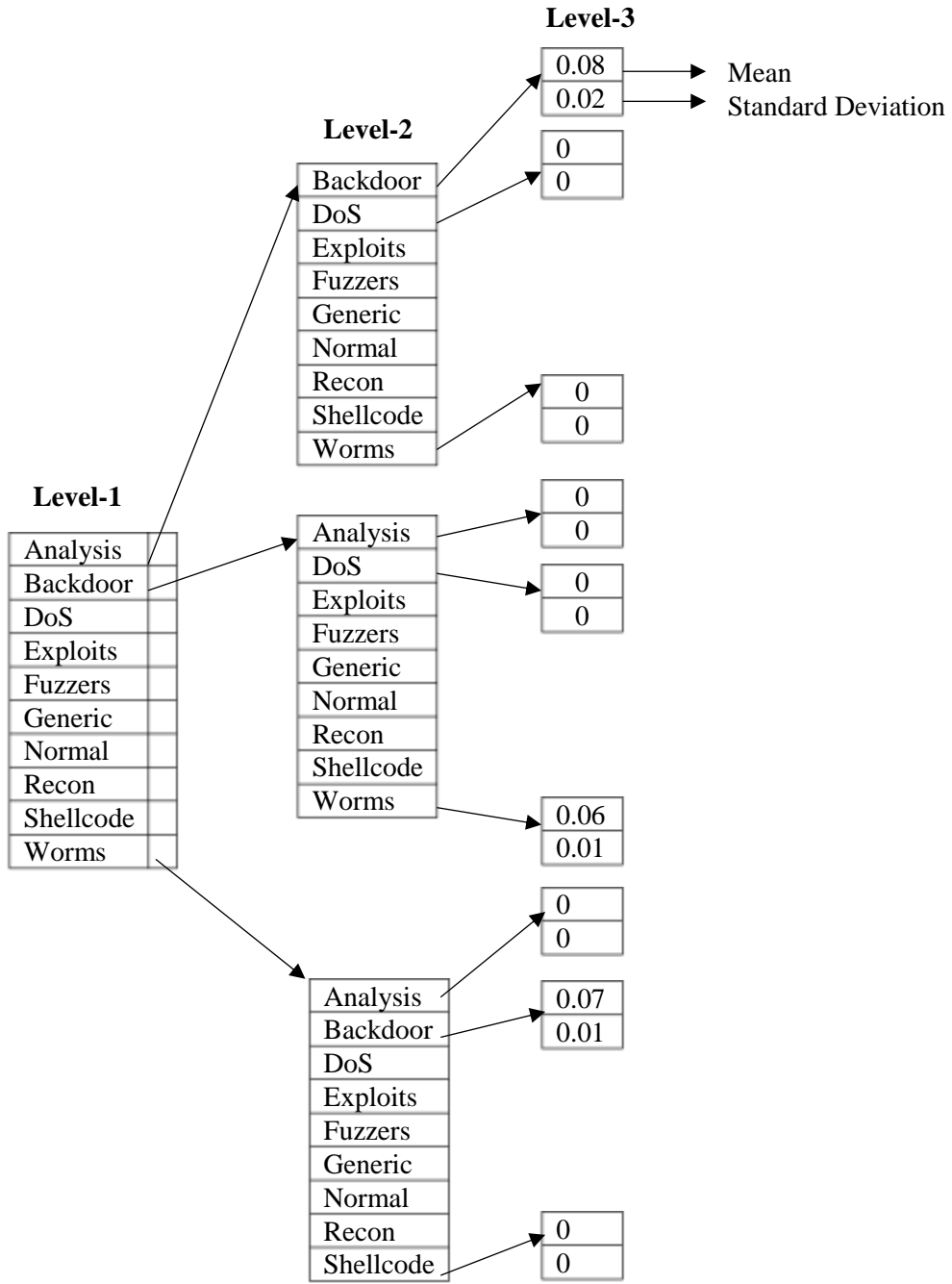

**Figure 3.** Illustration of partial output by Algorithm #1

Next, consider the initialization steps in lines 1 through line 3 for the very first iteration of this algorithm:

- Analysis class is removed from the class list held in AL and placed in variable $x$ as in line 1.

- In line 2, the Backdoor attack class is placed in variable $y$.

- Starting in line 3 and to compute the elements of both mean and standard deviation arrays, Algorithm #1 counts the false negatives for which Analysis is the target class while Backdoor is mistaken as Analysis due to misclassification. For this example, 124 records of Analysis are misclassified as Backdoor. Since the false-negative count is greater than zero for the Backdoor attack class, the difference between its prediction scores and those of Analysis are calculated for the corresponding data points. For instance, in the case of Analysis and Backdoor attack classes, we have 124 rows and 10 columns of probability scores generated by the BB classifier.

Execution now reaches the for-loop in line 5 during the first iteration.

- In line 5, the algorithm starts from the first row of the probability score matrix in which 0.78, 0.81, 0.02, 0.01, 0.24, 0.08, 0.11, 0.19, 0.08, 0.22 are the likelihood of a new Analysis attack sample being classified as Analysis, Backdoor, DoS, Exploits, Fuzzers, Generic, Normal, Reconnaissance, Shellcode, and Worms, respectively.

- In line 6, the differences between the Backdoor score and the scores of other attack classes are calculated.

- In line 6, since the difference when the score for Analysis (the target class with a score of 0.78) which is deducted from the score for Backdoor (incorrectly predicted class with a score of 0.81) is less than other differences, i.e. (0.81-0.78=0.03) < (0.81-0.02= 0.79) and so on, the value 0.03 is stored in the distance variable D.

- Difference value stored in D is transferred to the array DL in line 14.

The for-loop in line 5 repeats for 123 times. In the second iteration, the algorithm accesses the second row of the probability scores matrix with the values of 0.09, 0.54, 0.01, 0.03, 0.41, 0.04, 0.01, 0.06, 0.12 and 0.14.

- Line 6, calculates the differences of the scores.
- Since the difference between the Backdoor score and the Analysis score (0.54-0.09=0.45) is not less than the difference between Backdoor score and Fuzzers score (0.54-0.41=0.13), the difference will not be stored in the difference variable D. This computation repeats for the other 122 data points.
- In lines 22 through 25, the mean and the standard deviation for all those values in distance array DL are calculated and kept in the mean and standard deviation arrays, namely OLM and OLSD, respectively, for the Analysis and in the elements belonging to Backdoor. Now, we have the range of values for which the errors due to data overlap will need to be minimized.

In the next step, we find whether the range of values calculated for each attack class overlaps or not. If the ranges overlap, the classifier will have difficulty distinguishing between the two overlapping classes. Addressing the overlap will help minimize the confusion for the ensemble classifier when making predictions for the test dataset. That is, if the calculated ranges of two classes overlap, the range of a class with the larger false negative count will be kept and the range of a class with the smaller false negative count will be changed to 0. For example, if the range of values for Backdoor and Analysis, and Backdoor and Worms overlap, and if Analysis data are misclassified as Backdoor for 124 data points and Worms data are misclassified as Backdoor for 2 data points, the mean and standard deviation calculated for Backdoor and Analysis will be kept in the list, while the mean and standard deviation will be changed to zero for the case where Worms attack records are misclassified as Backdoor. In other words, assume *x* has a value in the range of $[\mu_x - \sigma_x, \mu_x + \sigma_x]$ where the records in class A are misclassified as class B, and *y* assumes a value in the range of $[\mu_y - \sigma_y, \mu_y + \sigma_y]$ where the records in class C are incorrectly identified as class B; the ranges for *x* and *y* overlap:

If $\quad\quad\quad\quad\quad\quad\quad \mu_y \in [\mu_x - \sigma_x, \mu_x + \sigma_x]$

Else if $\quad\quad\quad\quad\quad (\mu_y + \sigma_y) \leq (\mu_x + \sigma_x) \text{ and } (\mu_y - \sigma_y) \geq (\mu_x - \sigma_x)$

Else if $\quad\quad\quad\quad\quad (\mu_y - \sigma_y) \geq (\mu_x - \sigma_x) \text{ and } (\mu_y + \sigma_y) \leq (\mu_x + \sigma_x)$

In this case, the values of mean and standard deviation for *y* will remain unchanged if and only if the number of records in class A that are incorrectly classified as class B is greater than the number of records in class C that are misidentified as class B and then the values of mean and standard deviation for *x* will be changed to zero.

Figure 4 shows that the Fuzzers, Backdoor, and Normal classes overlap. In this figure, Fuzzers is associated with a range of values between 0.09-0.02=0.07 and 0.09+0.02=0.11; Backdoor has a range of values between 0.08-0.02=0.06 and 0.08+0.02=0.1; and Normal is associated with a range of values between 0.04-0.03=0.01 and 0.04+0.03=0.07, where Analysis is incorrectly predicted as Fuzzers, Backdoor, and Normal, respectively. These values are taken from Table 7 (b) and after finding the overlaps, Table 7 (c) would have the final mean and standard deviation values generated for all the classes. In Table 7 (c) the mean and standard deviation values for the Recon class remains unchanged since its error range is from 1.3-0.1=1.2 to 1.3+0.1=1.4 and does not have any overlap with other ranges. Since Fuzzers, Backdoor, and Normal ranges overlap as well as the greater number of Analysis records are misclassified as Backdoor, the mean and standard deviation values calculated for Backdoor remain unaltered, and the mean and standard deviation values associated with Fuzzers and Normal are changed to zero.

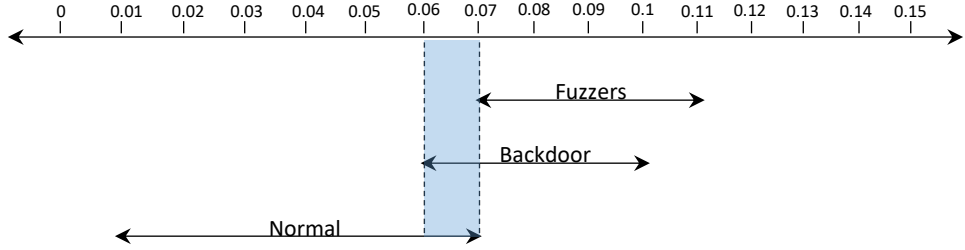

**Figure 4.** An illustration for the prediction error ranges for three classes for Algorithm #1

**Table 7.** (a) The confusion matrix, (b) Mean and standard deviation (SD) arrays through Algorithm #1 and (c) Mean and standard deviation (SD) arrays after the range revision

|  |  | Analysis | Backdoor | DoS | Exploits | Fuzzers | Generic | Normal | Recon | Shellcode | Worms |
|---|---|---|---|---|---|---|---|---|---|---|---|
| Analysis |  | 501 | 124 | 0 | 0 | 46 | 0 | 3 | 3 | 0 | 0 |

(a)

|  |  | Analysis | Backdoor | DoS | Exploits | Fuzzers | Generic | Normal | Recon | Shellcode | Worms |
|---|---|---|---|---|---|---|---|---|---|---|---|
| Analysis | Mean |  | 0.08 | 0 | 0 | 0.09 | 0 | 0.04 | 1.30 | 0 | 0 |
|  | SD |  | 0.02 | 0 | 0 | 0.02 | 0 | 0.03 | 0.10 | 0 | 0 |

(b)

|  |  | Analysis | Backdoor | DoS | Exploits | Fuzzers | Generic | Normal | Recon | Shellcode | Worms |
|---|---|---|---|---|---|---|---|---|---|---|---|
| Analysis | Mean |  | 0.08 | 0 | 0 | 0 | 0 | 0 | 1.30 | 0 | 0 |
|  | SD |  | 0.02 | 0 | 0 | 0 | 0 | 0 | 0.10 | 0 | 0 |

(c)

The process of finding the range overlaps and addressing them takes place for all 10 classes. This procedure generates a list of mean and standard deviation values as shown in Figure 3. The list is eventually utilized in the test phase to minimize the prediction errors as a result of class overlap.

## 4.2. Testing

In this phase, mean and standard deviation values computed by Algorithm #1 and revised after computing the class data overlaps are used to reduce the errors made by XGBoost and BB classifiers in identifying unseen data for class membership. Due to class data overlap associated with the UNSW-NB15, a new sample mimicking the behavior of data points belonging to another attack class is most likely to be misclassified. The probability scores matrix generated by the classifiers for a new sample in class A contains an error if this sample is incorrectly classified as class B. In this case, the probability matrix score for class

A is lower than that of class B, while it must be just the opposite. We consider the difference between the probability scores for classes A and B as the error value. To reduce this error, we apply Algorithm #2, presented in Figure 5, on the probability score matrices generated for XGBoost and BB classifiers.

The test set is utilized to evaluate the performance of the proposed design, and therefore, the classifier does not know the value of the real target variable and we cannot calculate the errors using the ground truth. Algorithm #2 is designed to go through the membership scores generated by XGBoost and BB classifiers during the training and validation phases in order to calculate the errors for the test subset regardless of values for the real target variables. The following steps elaborate on the functionality of Algorithm #2:

- In lines 3 to 8, the maximum (class) membership score using the probability scores matrix for each testing sample is found among ten different scores generated by a particular classifier, namely one of XGBoost and BB. This membership score is stored in the temporary variable max and its index value is kept in the temporary variable index_max.
- The algorithm calculates the difference values of max with nine other membership scores in lines 10 through 15. It computes the differences of probability score matrix values for other classes residing across the columns of probability scores matrix with the variable max which represents the maximum class membership score computed in lines 3 through 8. It picks the smallest such difference to store in variables min and index_min, which represent the difference value and class membership, respectively. These lines represent the computations to find the membership score corresponding to a class which is the second likely class.
- In lines 16 and 17, the corresponding probability score matrix value is modified using the associated mean value. Variables index_min and index_max represent the indices for two matrices OLM and OLSD and refer to the attack classes with the same ordering as in the list AL:

for instance, index_min = 1 means Analysis. The operation in Line 17 increases the likelihood of a sample becoming a member of the correct class type by adding the corresponding mean value calculated by Algorithm #1.

Using the mean and standard deviation values calculated by Algorithm #1, Algorithm #2 can mitigate the errors in the probability matrices where the errors arise due to class data overlap. Nevertheless, it must be noted that although errors are not zeroed out using this algorithm, they are reduced noticeably.

---

Algorithm #2 Membership Score Modification

---

Membership-Score-Modification (PS, OLM, OLSD)
    in:
        two-dimensional array $PS_{n \times 10}$ holding the membership scores, n = the number of samples in the test subset
        two-dimensional array $OLM_{10 \times 10}$
        two-dimensional array $OLSD_{10 \times 10}$
    out:
        two-dimensional array $PS_{n \times 10}$ holding the (modified) membership scores, n = the number of samples in the test subset

1: **for** i ← 1 … n **do**
2:   max ← 0   ▷ holding zero in max to find the maximum membership score from line 3 to 8
3:   **for** j ← 1 … 10 **do**
4:     **if** $PS_{i,j}$ > max **then**
5:       max ← $PS_{i,j}$
6:       index_max ← j
7:     **end if**
8:   **end for**
9:   min ← $10^{10}$   ▷ holding a very big constant value in min to find the minimum values obtained thru lines 10 to 15
10:   **for** j ← 1 … 10 **do**
11:     **if** $\left((\text{max} - PS_{i,j}) < \text{min}\right)$ **and** (index_max ≠ j) **then**
12:       min ← max − $PS_{i,j}$
13:       index_min ← j
14:     **end if**
15:   **end for**
16:   **if** (min ≥ ($OLM_{index\_min,index\_max} - OLSD_{index\_min,index\_max}$)) **and** (min ≤ ($OLM_{index\_min,index\_max} + OLSD_{index\_min,index\_max}$)) **then**
17:     $PS_{i,index\_min} \leftarrow (PS_{i,index\_min} + OLM_{index\_min,index\_max})$
18:   **end if**
19: **end for**
20: **return** PS

---

**Figure 5**. Pseudocode for Algorithm #2

The modified membership scores were used along with the membership scores obtained by the augmented RF in the testing phase to make the final prediction using the majority vote. Since hard voting is utilized in this method, the membership scores measured by each model for a particular class label are added and the class with the highest score is predicted in the majority vote. Overall classifier development steps are presented in Figure 6.

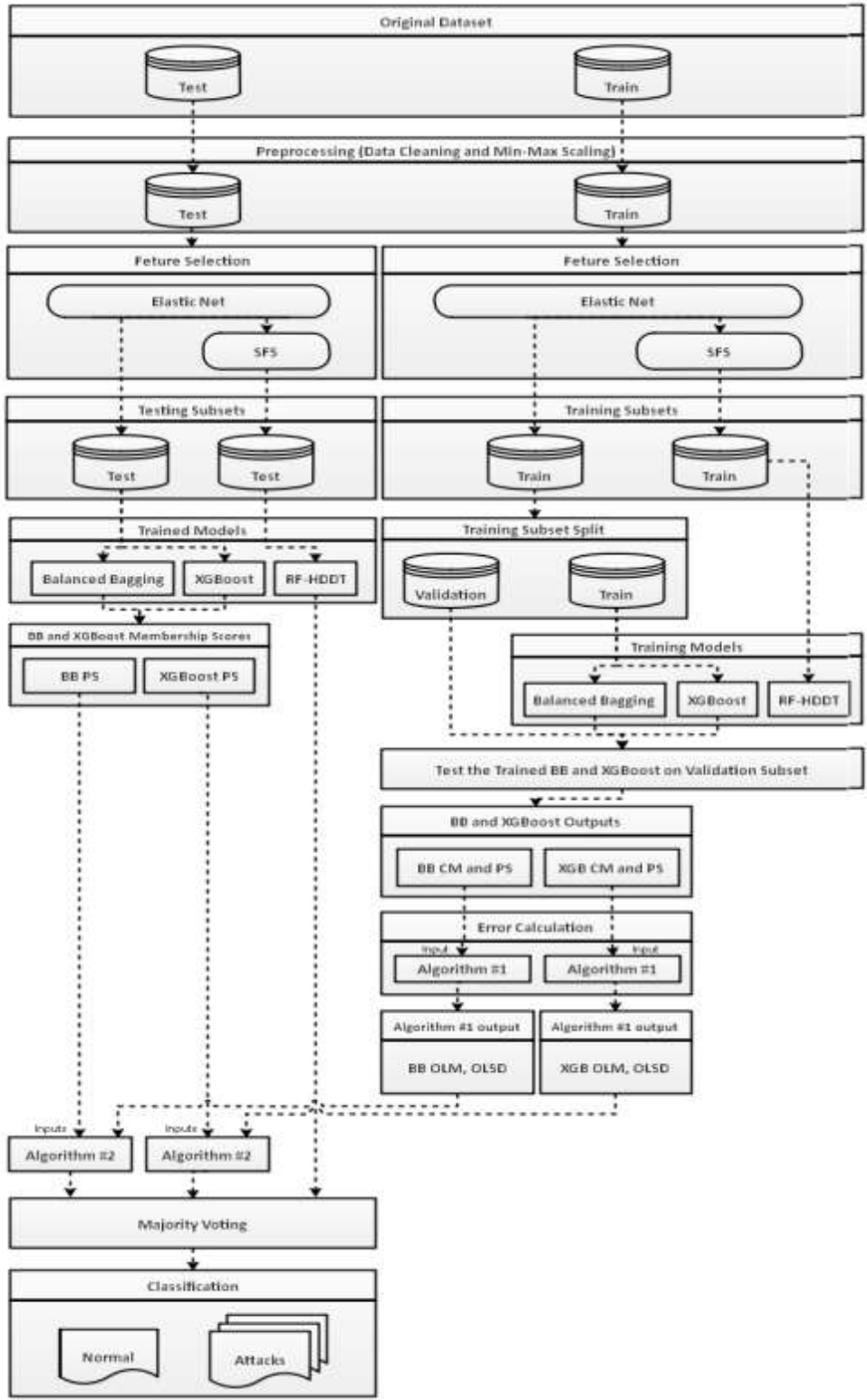

**Figure 6.** Classifier development process schematic diagram

# 5. Simulation Results

In this section, the proposed ensemble classifier is implemented and its performance on the UNSW-NB15 dataset is evaluated for both binary and multi-category classification. The results are compared with those published by others in the literature for the UNSW-NB15 dataset. The performance of each machine learning classifier is profiled using sensitivity, specificity, false positive rate, false negative rate, precision, and f-measure metrics for the comparison.

## 5.1. Binary Classification

Performance of the ensemble classifier on the test set with binary class labels, attack vs. Normal, is presented in Table 8 that shows 790 out of 45,332 attack records are misclassified. It means less than 2% of the attacks are misclassified as non-attack or Normal which indicates a missed alarm rate of 0.017 while the sensitivity for the attack class is 98.26%. On the other hand, 1,022 of 37,000 Normal records are incorrectly classified as attack which indicates less than 3% false alarm rate. Several evaluation metric values are shown in Table 9 in order to comprehensively assess the performance of the proposed classifier design.

**Table 8.** Confusion Matrix for test data for binary classification

| Predicted / Actual | Normal | Attack |
|---|---|---|
| Normal | 35978 | 1022 |
| Attack | 790 | 44542 |

**Table 9.** Performance of the proposed design for binary classification

| Metrics / Class | Sensitivity (%) | Specificity (%) | False Positive Rate | False Negative Rate | Precision (%) | F-measure (%) |
|---|---|---|---|---|---|---|
| Normal | 97.24 | 98.26 | 0.017 | 0.028 | 97.85 | 97.75 |
| Attack | 98.26 | 97.24 | 0.028 | 0.017 | 97.76 | 98.01 |

The main objective of any intrusion detection system (IDS) is to identify the pattern of the network traffic that may imply a suspicious activity. Accordingly, performance of the proposed IDS on UNSW-NB15 data is competitive in comparison with the other studies reported in the literature as shown in Table 10. The second column from the left presents the values of performance metrics associated with the attack class in Table 9, which suggests that the performance of the proposed classifier design is superior in comparison to all those listed in Table 10, namely studies reported in [12, 46-50].

**Table 10.** Comparison of the proposed classifier design with four others cited (NR:Not Reported)

| Metrics | Proposed Design | [12] | [46] | [47] | [48] | [49] | [50] |
| --- | --- | --- | --- | --- | --- | --- | --- |
| Sensitivity (%) | 98.26 | 85.00 | NR | 98.47 | 91.30 | 92.10 | 99.74 |
| FNR (%) | 01.74 | 15.00 | NR | NR | NR | 08.10 | 1.31 |
| FPR (%) | 02.76 | 02.00 | 02.97 | 02.18 | 08.90 | 05.80 | NR |
| Precision (%) | 97.76 | 99.00 | NR | NR | 91.60 | 94.10 | 91.66 |
| F-measure (%) | 98.01 | 91.00 | NR | NR | NR | 93.08 | 95.53 |
| Accuracy (%) | 97.80 | 89.00 | 95.08 | 94.11 | 91.27 | 92.20 | 98.82 |

The authors in [12] implemented MLP as an anomaly detection system for binary classification. They employed Recursive Feature Elimination algorithm along with the RF classifier for the purpose of dimensionality reduction. This method selected the top four informative features. The original training and testing subsets proposed by [27], are used for training and evaluating the model. The MLP-based IDS scored 85% for sensitivity and 89% for accuracy on the test subset of UNSW-NB15: 15% of the attack traces and 2% of the normal records were misclassified.

Bayu et al. [46] applied Gradient Boosted Machine (GBM) on three datasets including UNSW-NB15. No feature selection technique was implemented. All 47 features were kept for training and testing phases. The results showed that GBM outperformed four other algorithms, namely RF, Deep Neural Network (DNN), SVM and CART, with the average accuracy value of 93.64% and missed alarm rate of 0.0206 where GBM performance was evaluated on NSL-KDD, UNSW-NB15 and GPRS datasets. While running GBM on UNSW-NB15 alone provided 95.08% accuracy and 2.97% false alarm rate using 10-fold cross-validation and the accuracy of 91.31% and false alarm rate of 8.60% using the Hold-Out method on the original UNSW-NB15 train and test subsets.

In [47], nominal features were converted to numerical and then the Min-Max normalization method was utilized to scale down the values to the range of 0 to 1. They used 5-fold cross-validation without resampling to generate the test and training subsets. They calculated the average values for the sensitivity, false alarm rate and accuracy of 5 folds. The authors utilized SVM by taking advantage of hyper clique property of hypergraph to improve the performance of SVM. This optimization technique implemented the feature selection as well. The SVM algorithm was trained with the entire 47 features and then the results were compared with the case when SVM was trained with the optimal number of feature subsets. The optimal feature subset is not reported. However, the number of optimal features is in the range of 30 to 35. They concluded that the feature selection had significant influence on the proposed model which delivered 98.47% sensitivity and 2.18% false alarm rate. However, 94.11% accuracy and 2.18% false alarm rate suggests a relatively large value for the missed alarm rate, which is not reported.

In [48] a hybrid feature selector consisting of particle swarm optimization, ant-colony algorithm, and genetic algorithm was used to reduce the dimensionality of the training dataset. The performance of the reduced error pruning tree (REPT) was evaluated to determine the most informative features in the feature selection phase. A two-level ensemble based classifier comprising rotation forest and bagging classifier was implemented on the NSL-KDD and the UNDW-NB15 datasets. The classifier yielded 91.30% sensitivity while classifying 91.60% of the positive samples correctly.

Researchers in [49] proposed the parallel K-medoids algorithm using map reduction function to cluster the instances. In this study, 8.10% of the malicious activities were misclassified while 5.8% of the normal class members were classified as belonging to attack classes.

Zhang, et al. [51] proposed the NIDS method consisting of a data preprocessing module, an imbalance processing module, and a classification decision module, namely SGM-CNN. The performance of the proposed model was evaluated on the UNSW-NB15 and CICIDS2017 datasets in terms of binary classification and multi-category classification. Their design achieved 99.74% detection rate as a binary classifier.

## 5.2. Multiclass Classification

Performance evaluation of the ensemble classifier (without the application of the proposed algorithms #1 and #2) for the multiclass case is shown in Table 11 in terms of the confusion matrix. Although the ensemble design has effectively improved the performance of each base classifier in comparison, it still suffered from generating a relatively large number of false negatives. The two algorithms, namely Algorithms #1 and #2, were proposed specifically to address this issue: the resulting ensemble classifier was then named as the proposed ensemble classifier or design.

**Table 11.** Confusion matrix showing the performance of the ensemble classifier

|           | Analysis | Backdoor | DoS | Exploits | Fuzzers | Generic | Normal | Recon | Shellcode | Worms |
|-----------|----------|----------|-----|----------|---------|---------|--------|-------|-----------|-------|
| Analysis  | 327      | 193      | 4   | 1        | 18      | 0       | 130    | 0     | 4         | 0     |
| Backdoor  | 298      | 154      | 8   | 8        | 23      | 0       | 76     | 1     | 15        | 0     |
| DoS       | 1477     | 737      | 775 | 435      | 221     | 0       | 228    | 45    | 164       | 7     |
| Exploits  | 1322     | 1583     | 51  | 6197     | 257     | 0       | 667    | 490   | 420       | 145   |
| Fuzzers   | 651      | 386      | 8   | 14       | 1837    | 0       | 2604   | 7     | 540       | 15    |
| Generic   | 15       | 30       | 37  | 390      | 99      | 18145   | 63     | 7     | 72        | 13    |
| Normal    | 842      | 1        | 3   | 28       | 1427    | 0       | 34545  | 0     | 154       | 0     |
| Recon     | 97       | 210      | 0   | 35       | 14      | 0       | 31     | 2967  | 128       | 14    |
| Shellcode | 11       | 0        | 0   | 0        | 7       | 0       | 9      | 3     | 347       | 1     |
| Worms     | 0        | 0        | 0   | 1        | 0       | 0       | 1      | 0     | 3         | 39    |

The performance of the proposed ensemble classifier with the algorithms #1 and #2 incorporated is shown in Table 12. Comparing results in Table 11 with those in Table 12 indicates a significant improvement for most classes including the Normal class which incidentally results in a notable reduction in the number of false negatives. This improvement verifies the effectiveness and utility of algorithms #1 and #2.

**Table 12.** Confusion matrix showing the performance of the proposed design

|  | Analysis | Backdoor | DoS | Exploits | Fuzzers | Generic | Normal | Recon | Shellcode | Worms |
|---|---|---|---|---|---|---|---|---|---|---|
| Analysis | 327 | 193 | 4 | 1 | 18 | 0 | 130 | 0 | 4 | 0 |
| Backdoor | 126 | 405 | 0 | 0 | 13 | 0 | 39 | 0 | 0 | 0 |
| DoS | 1604 | 625 | 1110 | 228 | 92 | 0 | 178 | 59 | 164 | 29 |
| Exploits | 779 | 830 | 39 | 8511 | 57 | 0 | 338 | 221 | 213 | 144 |
| Fuzzers | 482 | 491 | 4 | 34 | 4380 | 0 | 0 | 10 | 646 | 15 |
| Generic | 15 | 30 | 37 | 390 | 99 | 18145 | 63 | 7 | 72 | 13 |
| Normal | 200 | 0 | 0 | 0 | 668 | 0 | 35978 | 0 | 154 | 0 |
| Recon | 103 | 207 | 0 | 35 | 14 | 0 | 31 | 2967 | 126 | 13 |
| Shellcode | 0 | 0 | 0 | 0 | 7 | 0 | 9 | 3 | 358 | 1 |
| Worms | 0 | 0 | 0 | 1 | 0 | 0 | 2 | 0 | 5 | 36 |

In Table 12, the number of attack records misclassified as Normal is reduced significantly compared to the same column in Table 11 indicating that the overall detection rate of attack class records improved noticeably. Comparing the entries along the diagonals of Tables 11 and 12 further shows that the detection rate for a given attack class membership also improved for most attack classes. There are many nonzero off-diagonal entries in the confusion matrix in Table 12 indicating that the classifier still cannot distinguish among the attack classes for a non-negligible number of attack records.

The performance of the proposed ensemble classifier is presented in terms of benchmark metrics in Table 13. As shown in Tables 12 and 13, the share of attack records that are incorrectly categorized as Normal is 4.14% out of the 28% overall missed alarm rate. The remaining 23.86% missed alarm rate is associated with misclassifications among attack classes which may not be construed as problematic for intrusion detection purposes as this indicates that an attack was detected but the type of attack is not identified correctly. Although 4.14% missed alarm rate for attack records alone could be detrimental for an intrusion detection system, our design achieves better results in comparison with the classifiers in other studies as shown in Table 14.

Normal traces are also misclassified as Shellcode, Fuzzers and Analysis which suggests approximately 3% false alarm rate and 97.24% sensitivity as shown in Table 13. This is because these attack types mimic the behavior of Normal records [30, 52, 53], which is the main reason that some attacks are traced incorrectly and are predicted as Normal activity. In the associated confusion matrix in Table 12, 19.20% of Analysis, 6.69% of Backdoor, 4.35% of DoS, 3.04% of Exploits, 0.33% of Generic, 0.88% of Reconnaissance, 2.38% of Shellcode, and 4.55% of Worms attack records are confused with Normal records.

**Table 13.** Performance of the proposed design for the multi-class case

| Metric / Class | Accuracy | Sensitivity | Specificity | FPR | FNR | Precision | F-measure |
|---|---|---|---|---|---|---|---|
| Analysis | 95.56 | 48.30 | 95.95 | 0.041 | 0.51 | 0.15 | 64.25 |
| Backdoor | 96.89 | 69.47 | 97.09 | 0.029 | 0.31 | 0.84 | 80.99 |
| Dos | 96.27 | 27.15 | 99.89 | 0.001 | 0.73 | 0.42 | 42.70 |
| Exploits | 95.98 | 76.46 | 99.03 | 0.009 | 0.24 | 0.84 | 86.29 |
| Fuzzers | 96.78 | 72.25 | 98.73 | 0.012 | 0.28 | 0.77 | 83.44 |
| Generic | 99.12 | 96.15 | 100.00 | 0.000 | 0.28 | 1.00 | 98.04 |
| Normal | 97.81 | 97.24 | 98.26 | 0.017 | 0.02 | 0.97 | 97.75 |
| Reconnaissance | 95.39 | 84.87 | 95.86 | 0.041 | 0.15 | 0.61 | 90.03 |
| Shellcode | 99.31 | 94.71 | 98.33 | 0.017 | 0.05 | 0.34 | 96.49 |
| Worms | 99.73 | 81.82 | 99.74 | 0.003 | 0.18 | 0.24 | 89.90 |

Performance comparison of the proposed model with those studies reported in the literature is presented in Tables 14 and 15. Many of the relevant performance metrics including the missed alarm rate, which is one of the most critical ones for the intrusion detection context, are not reported in these studies by others. Consequently, a performance comparison is done only for accuracy and sensitivity as these are the only metrics commonly reported in the cited studies.

**Table 14.** Accuracy of the proposed design vs. the accuracy of different classifier (NR: Not Reported)

| Class | Accuracy (Proposed) | Accuracy [54] | Accuracy [55] | Accuracy [56] | Difference |
|---|---|---|---|---|---|
| Analysis | 95.56 | 99.26 | **99.30** | NR | **-3.74** |
| Backdoor | 96.89 | **99.11** | 97.93 | NR | **-2.22** |
| Dos | **96.27** | 94.90 | 95.71 | 94.52 | +0.56 |
| Exploits | **95.98** | 90.12 | 93.58 | 89.72 | +2.40 |
| Fuzzers | **96.78** | 91.47 | 95.04 | NR | +1.74 |
| Generic | **99.12** | 98.23 | 98.70 | 87.70 | +0.42 |
| Normal | 97.81 | 93.54 | 94.59 | **98.64** | **-0.30** |
| Reconnaissance | 95.39 | 95.33 | 96.18 | **99.10** | **-3.71** |
| Shellcode | 99.31 | **99.40** | 98.33 | NR | **-0.09** |
| Worms | 99.73 | **99.92** | 99.78 | NR | **-0.14** |

**Table 15.** Sensitivity of the proposed model vs. the sensitivity of different classifiers (NR: Not Reported)

| Attack Type | Sensitivity (Proposed) | Sensitivity [55] | Sensitivity [56] | Sensitivity [51] | Difference |
|---|---|---|---|---|---|
| Analysis | **48.30** | 20.45 | NR | NR | +27.85 |
| Backdoor | **69.47** | 67.32 | NR | NR | +02.15 |
| Dos | **27.15** | 14.29 | 5.0 | 29.46 | +12.86 |
| Exploits | **76.46** | 76.22 | 54.64 | 85.69 | +00.24 |
| Fuzzers | **72.25** | 64.42 | NR | NR | +07.83 |
| Generic | 96.15 | 81.37 | **96.72** | 98.58 | **-00.57** |
| Normal | 97.24 | 97.39 | **98.00** | 96.19 | **-00.76** |
| Reconnaissance | **84.87** | 46.04 | 71.70 | 79.42 | +13.17 |
| Shellcode | **94.71** | 36.39 | NR | NR | +58.32 |
| Worms | **81.82** | 18.37 | NR | NR | +63.45 |

Kumar et al. [51] developed the Unified Intrusion Detection System (UIDS) by generating the new training and test subsets out of UNSW-NB15. They utilized the k-means clustering algorithm to increase the attack sensitivity as the k-means clustering algorithm was able to identify the similarities between different attack classes. In each cluster, the number of records in some type of attack classes was more than the rest. The authors randomly picked 65% of the records of the dominating class categories to form a training dataset. The remaining 35% of the instances were used to build the test set. They also used information gain algorithm for the feature selection phase. 13 features out 47 were selected due to the improvement of accuracy scores by C5.0, Chi-Squared Automatic Inference Detection (CHAID), Classification and Regression Tree (CART) is also known as Decision Tree (DT) and Quick Unbiased Efficient Statistical Tree (QUEST) algorithms. These algorithms were used to form the proposed UIDS model. Their study reported 77.87% and 79.12% for the average sensitivity and F-measure, respectively. It also offered 3.80% false alarm rate for Normal instances and 86.15% attack sensitivity.

Papamartzivanos et al. [55] combined the decision tree and genetic algorithm to generate classification rules and called their model as Dendron. The wrapper technique was used for feature selection, which resulted in 23 being selected as informative features. The authors reported a sensitivity of 97.39% for the Normal class and the average false alarm rate of 2.61% as presented in Table 15. In this study, 10% of the instances for each of the 9 classes were considered for building the training set and the remaining 90% was kept for the test set. To address the imbalance problem of multi-class classification in the UNSW-NB15 data set, 50% of Worms attack class records were included in the training subset, and the remaining 50% was kept to test the model. The proposed model in our study outperforms the other two given the F-measure values. The likely explanation is that the imbalance and overlapping problems for the proposed ensemble classifier are addressed with some degree of success.

The integrated rule-based classifier is trained to detect five class types to avoid overlapping [45]. The rules employed by this classifier were generated from four tree-structured classification algorithms, namely C5.0, CHAID, CART, and QUEST. Training and test sets were built by eliminating some instances from the original training and testing subsets using k-means clustering. These eliminated instances belong to Analysis, Backdoor, Fuzzers, Shellcode, and Worms attack types as these attack types suffer from the class overlap problem and their presence may cause poor classification performance. For the feature selection phase, the genetic algorithm was used and 22 features were included. They reported good accuracy and sensitivity values for the Normal class. Also, the reconnaissance is successfully detected with an accuracy of 99.10% as shown in Table 15. However, the average accuracy and sensitivity values are 93.94% and 65.21%, respectively.

Figure 7 depicts the performance of the proposed ensemble design in comparison with two other models, namely Integrated [56] and Dendron [55] in terms of the F-measure. As this figure indicates the proposed ensemble classifier outperforms both Dendron and Integrated for each and every class.

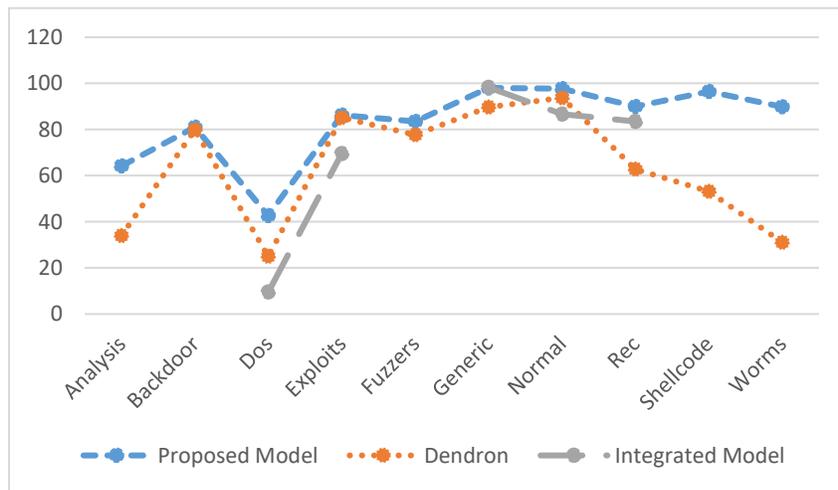

**Figure 7.** Comparison of the proposed design with two others in literature based on the F-measure

# 6. Conclusions

This study presented the design, development, and performance evaluation of an ensemble classifier that was tuned to the characteristics of a cybersecurity dataset, namely UNSW-NB15, for both multi-category and binary classification cases. A combination of preprocessing [57], feature selection, tree-based base classifiers, ensemble classification framework, and two novel algorithms were proposed to address the inherent issues with the dataset. The performance of the proposed classifier was compared with both the binary classifiers and multi-category classifiers reported in the literature on the same dataset.

In the binary classification case, the proposed ensemble design could classify more than 98% of the attack class records correctly leading to less than 2% of the attack class records being misclassified. Its performance for the classification of Normal records indicated that more than 97% were classified correctly with less than 3% false alarm rate. For the multi-class case, 28% of the attack records were misclassified: 4.14% of the attack records were mislabeled as belonging to the Normal class, and the remaining 23.86% of attack class records were misclassified as belonging to other attack classes. Additionally, 97.24% of Normal traces are classified correctly and the remaining 2.76% were due to false alarms as they were misclassified as Shellcode, Fuzzers, and Analysis. Almost 100% of Generic and Reconnaissance class records were classified correctly. In terms of missed alarm cases, 19.20% of Analysis, 6.69% of Backdoor, 4.55% of Worms, 4.35% of DoS, 3.04% of Exploits, and 2.38% of Shellcode attack records were incorrectly labeled as Normal records.

In comparison with other studies reported in the current literature on the UNSW-NB15 dataset, the proposed ensemble design achieved significantly improved performance. The main reason for this performance improvement was that the proposed ensemble design addresses two major issues that this dataset suffers from, namely data overlap and class imbalance. The proposed algorithms minimize the errors caused by data overlap and class imbalance. These algorithms are generic and therefore can be used with any other machine learning classifiers that employ tree-based algorithms.

40. Kasongo, S.M. and Y. Sun, *Performance analysis of intrusion detection systems using a feature selection method on the UNSW-NB15 dataset.* Journal of Big Data, 2020. **7**(1): p. 1-20.

41. Zhang, J., et al., *Model of the intrusion detection system based on the integration of spatial-temporal features.* Computers & Security, 2020. **89**: p. 101681.

42. Kasongo, S.M. and Y. Sun, *A deep learning method with wrapper based feature extraction for wireless intrusion detection system.* Computers & Security, 2020. **92**: p. 101752.

43. Barandela, R., R.M. Valdovinos, and J.S. Sánchez, *New applications of ensembles of classifiers.* Pattern Analysis & Applications, 2003. **6**(3): p. 245-256.

44. Aridas, C. *imbalanced-learn*. 2019 [cited 2019; Available from: https://github.com/scikit-learn-contrib/imbalanced-learn/.

45. Nguyen, T.D., et al., *Stratified random sampling from streaming and stored data.* Distributed and Parallel Databases, 2020: p. 1-46.

46. Tama, B.A. and K.-H. Rhee, *An in-depth experimental study of anomaly detection using gradient boosted machine.* Neural Computing and Applications, 2019. **31**(4): p. 955-965.

47. Gauthama Raman, M., et al., *An efficient intrusion detection technique based on support vector machine and improved binary gravitational search algorithm.* Artificial Intelligence Review, 2020. **53**(5): p. 3255-3286.

48. Tama, B.A., M. Comuzzi, and K.-H. Rhee, *TSE-IDS: A two-stage classifier ensemble for intelligent anomaly-based intrusion detection system.* IEEE Access, 2019. **7**: p. 94497-94507.

49. Dahiya, P. and D.K. Srivastava. *A comparative evolution of unsupervised techniques for effective network intrusion detection in hadoop*. in *International Conference on Advances in Computing and Data Sciences*. 2018. Springer.

50. Zhang, H., et al., *An effective convolutional neural network based on SMOTE and Gaussian mixture model for intrusion detection in imbalanced dataset.* Computer Networks, 2020. **177**: p. 107315.

51. Kumar, V., A.K. Das, and D. Sinha, *UIDS: a unified intrusion detection system for IoT environment.* Evolutionary Intelligence, 2021. **14**(1): p. 47-59.

52. Kumar, V., et al., *An integrated rule based intrusion detection system: analysis on UNSW-NB15 data set and the real time online dataset.* Cluster Computing, 2019: p. 1-22.

53. Yang, Y., et al., *Building an effective intrusion detection system using the modified density peak clustering algorithm and deep belief networks.* Applied Sciences, 2019. **9**(2): p. 238.
40

54. Sharma, J., et al., *Multi-layer intrusion detection system with ExtraTrees feature selection, extreme learning machine ensemble, and softmax aggregation.* EURASIP Journal on Information Security, 2019. **2019**(1): p. 1-16.

55. Papamartzivanos, D., F.G. Mármol, and G. Kambourakis, *Dendron: Genetic trees driven rule induction for network intrusion detection systems.* Future Generation Computer Systems, 2018. **79**: p. 558-574.

56. Kumar, V., et al., *An integrated rule based intrusion detection system: analysis on UNSW-NB15 data set and the real time online dataset.* Cluster Computing, 2020. **23**(2): p. 1397-1418.

57. Liu, J., et al., *Artificial Intelligence-Based Image Enhancement in PET Imaging: Noise Reduction and Resolution Enhancement.* PET clinics, 2021. **16**(4): p. 553-576.